\documentclass[aps,prb,twocolumn,amssymb,floatfix,epsfig]{revtex4}
\usepackage{graphicx,epsfig,psfrag,subfigure,amsopn}
\usepackage{color}
\usepackage{hyperref}
\usepackage{hypcap}
\usepackage{amssymb,amsbsy,amsmath}
\usepackage{amsmath} \textheight=22truecm
\newcommand\beq{\begin{equation}}
\newcommand\eeq{\end{equation}}
\newcommand\bea{\begin{eqnarray}}
\newcommand\eea{\end{eqnarray}}
\newcommand\bsq{\begin{subequations}}
\newcommand\esq{\end{subequations}}

\newcommand\bib{\bibitem}

\begin{document}

\title{Many-body localized phase of bosonic dipoles in a tilted optical lattice}

\author{Anirban Dutta$^1$, Subroto Mukerjee$^2$, and K. Sengupta$^3$}

\affiliation{$^1$Centre for High Energy Physics, Indian Institute of Science,
    Bengaluru 560012, India\\
    $^2$Department of Physics, Indian Institute of Science,
Bengaluru 560012, India \\
$^3$ Theoretical Physics Department, Indian Association for the
Cultivation of Science, Jadavpur, Kolkata 700032, India}

\date{\today}

\begin{abstract}

We chart out the ground state phase diagram and demonstrate the
presence of a many-body localized (MBL) phase for an experimentally
realizable  one-dimensional (1D) constrained dipole boson model in
the presence of an Aubry-Andre (AA) potential whose strength
$\lambda_0$ can be tuned to precipitate an ergodic-MBL transition.
We discuss the signature of such a transition in the quantum
dynamics of the model by computing its response subsequent to a
sudden quench of $\lambda_0$. We show that the MBL and the ergodic
phases can be clearly distinguished by study of post-quench dynamics
and provide an estimate for minimal time up to which experiments
need to track the response of the system to confirm the onset of the
MBL phase. We suggest experiments which can test our theory.

\end{abstract}
\maketitle

\begin{figure}[t!]
\begin{center}
\includegraphics[width=\columnwidth]{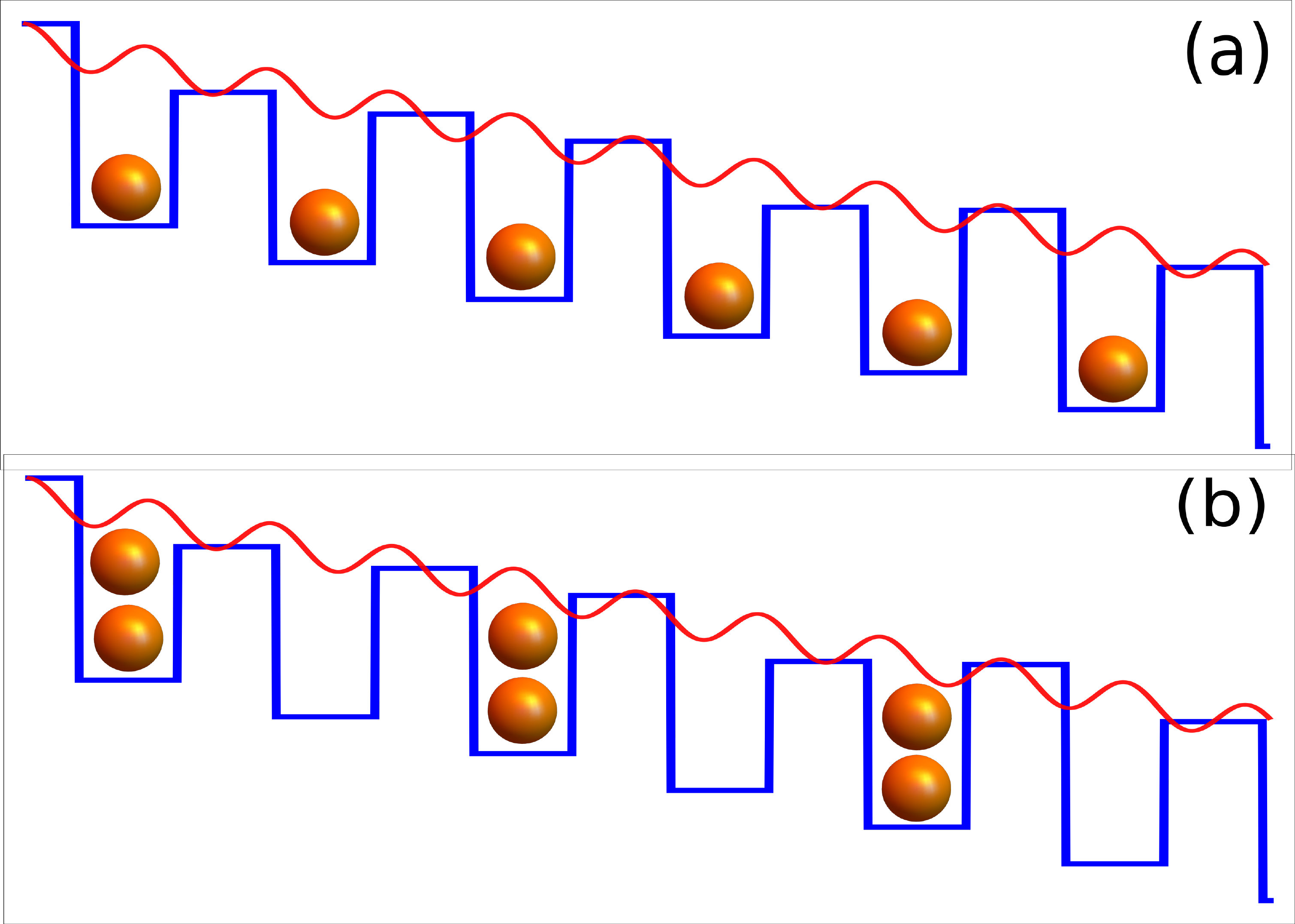}
\end{center}
\caption{(a) A schematic representation of the parent Mott state of
the tilted Bose-Hubbard model in the presence of the AA potential.
(b) A schematic representation of a $Z_2$ symmetry broken dipole
ordered state \cite{supp1}. } \label{fig1}
\end{figure}

Ultracold bosonic atoms in an optical lattice form one of the most
experimentally and theoretically well-studied strongly correlated
systems in recent times \cite{bloch1,rev1,tvr1, jaksch1,dupuis1}.
The central interest in these systems initially stemmed from the
experimental demonstration of the existence of a quantum phase
transition (QPT) of the constituent bosons from a superfluid to a
Mott insulating (MI) phase \cite{greiner1}. It was later realized
that in the presence of an effective electric field (or equivalently
in a tilted optical lattice), the bosons, in their MI state, undergo
yet another QPT which belongs to the Ising universality class
\cite{subir1}. Such an electric field can be generated either by
shifting the center of the trap used to confine the atoms
\cite{bloch1} or by applying a linearly varying Zeeman field
\cite{greiner1}. This transition takes the system from a parent MI
state with $n_0$ bosons per site to a $Z_2$ symmetry broken state
with $n_0 \pm 1$ bosons occupying every alternate site. The physics
of the transition and the phases separated by it is conveniently
described in terms of dipoles, which are bound states of bosons and
holes in adjacent sites as shown in Fig. \ref{fig1} \cite{subir1}.
The physics of these systems for $d>1$ \cite{subir1,subir2} and
their non-equilibrium dynamics for $d=1$ has also been studied
\cite{ks1,ks2}. The latter works \cite{ks1,ks2} indicated that such
systems can act as test beds for a realization of the Kibble-Zurek
mechanism. Furthermore, such constrained dipole models with
additional density-density interaction between bosons realize $Z_3$
and $Z_4$ symmetry broken phases\cite{subir3}; these models has
recently been experimentally realized using a Rydberg atom chain
\cite{greiner2}.

Many-Body Localization (MBL) in interacting quantum systems is one
of the most widely studied phenomena in recent times
\cite{mblrev1,mblrev2}. The loss of ergodicity in such systems due
to strong disorder or quasiperiodic potentials has been confirmed
theoretically by a wide variety of numerical and semi-analytic
studies \cite{mblpapersanalytical, mblpapersnumerical}. The
observation of MBL requires a high degree of isolation of the
experimental system from the environment, rendering systems of cold
atoms and ions as ideal testbeds. However, only a few such systems
are currently available \cite{ultrambl1,ionmbl1}. Thus,
identification of other, currently realizable, experimental systems
which may display MBL phases is of central importance to the field.

In this work, we show that the constrained dipole model realized
experimentally in Ref.\ \onlinecite{greiner1}, in the presence of an
additional Aubrey-Andre (AA) potential, supports an ergodic-MBL
transition (for the properties of the model in the presence of
on-site disorder, see Ref.\ \onlinecite{supp1}). The Hamiltonian of
this constrained dipole model can be described in terms of dipole
creation operator $d^{\dagger}_{\ell}$ on a link $\ell$ between two
sites $i$ and $j$ of the 1D lattice \cite{subir1}. These operators
are related to the creation ($b_j^{\dagger}$) and annihilation
($b_i$) operators of the original bosons: $d_l^{\dagger} = b_i
b_j^{\dagger}/\sqrt{n_0(n_0+1)}$. We analyze this dipole Hamiltonian
by carrying out exact diagonalization (ED) on finite-size boson
chains with $L\le 18$ to obtain our main results which are as
follows.

First, we compute the ratio of the difference of successive gaps in
the energy spectrum and the entanglement entropy, both of which can
be measured experimentally \cite{ultrambl1,ionmbl1}. In addition, we
compute the Normalized Participation Ratio (NPR)
\cite{mblpapersanalytical}. The behavior of all these quantities
demonstrate the existence of localized and ergodic phases in our
model and a transition between them. Second, we chart out the
long-time behavior of the dipole order parameter in the $Z_n$
symmetry broken phase (Fig.\ \ref{fig1}) where there is one dipole
every $n$ sites, $O_d^{(n)} =\sum_{\ell} d_{\ell}^{\dagger} d_{\ell}
\cos(2\pi \ell/n)/L$, for $n=3$ following a sudden quench of the AA
potential. The post-quench dynamics indicates thermalization (or
lack thereof) of $O_d^{(3)}$ in the ergodic (MBL) phase leading to
its qualitatively distinct behavior at long times in these two
phases. Third, we provide an estimate of the minimum time up to
which the experiments need to track the behavior of $O_d^{(3)}(t)$
to ascertain the onset of the MBL phase. We note that such
qualitatively distinct nature of $O_d^{(3)}(t)$ can be
experimentally detected via parity of occupation measurement
\cite{greiner1,greiner2}. This feature allows one to obtain an
unambiguous signature of the MBL phase within experimentally
relevant time scales. We also note that the Hamiltonian we study has
a truncated Hilbert space arising from the constraint of not
allowing dipoles to occupy adjacent sites, as will be explained
later. Our study therefore demonstrates that MBL can occur in
systems with truncated Hilbert spaces.

We begin by specifying the 1D boson Hamiltonian in an optical
lattice in the presence of an AA potential which is given by
\begin{eqnarray}
H_B &=& -J \sum_{\langle ij\rangle} b_i^{\dagger} b_j + \sum_j
[\frac{U}{2} n_j (n_j-1) + (\lambda_j-{\mathcal E} j) n_j] \nonumber
\label{ham0}
\end{eqnarray}
where $\lambda_j= \lambda_0 \cos(2\pi \beta j+\phi)$ is the AA
potential, $\beta=2/(\sqrt{5}-1)$ is the golden ratio conjugate,
${\mathcal E}$ is the effective electric field
\cite{greiner1,bloch1}, $U$ is the on-site interaction between the
bosons, $J$ is the hopping potential, $\phi$ is the offset angle,
and $\langle ij\rangle$ indicates that $j$ is one of the neighboring
sites of $i$. In the absence of $\lambda_j$ and in the regime $U,
{\mathcal E} \gg |U-{\mathcal E}|, J$, the low-energy physics of
this systems can be described in terms of dipole operators since
these excitations are resonantly connected to the parent Mott state
\cite{subir1}. The on-site energy for formation of these dipoles is
$\mu_d= U-{\mathcal E}$ (see Fig.\ \ref{fig1}) and the hopping term
$J$ allows for spontaneous creation and annihilation of these
dipoles leading to non-conservation of dipole number.

In the presence of the AA potential, the on-site energy cost for
creation of dipoles is modified. The bosons feel a difference in
potential originating from the AA term when it hops to the
neighboring site. A straightforward calculation yields
\begin{eqnarray}
\mu_d(\ell) &=& U -{\mathcal E} + V'_0 \cos(2\pi \beta \ell -\phi_0+\phi)
\label{onsiteeq}
\end{eqnarray}
where $V'_0=-2\lambda_0 \sin(\pi \beta)$ and $\phi_0=\pi/2$, and
$\phi$ may be used to produce different realizations of this
quasiperiodic potential. Thus the dipoles see an effective AA
potential with modified amplitude which can be controlled by tuning
$\lambda_0$. Moreover, as long we restrict ourselves to the regime
$U, {\mathcal E} \gg |U-{\mathcal E}|, J, \lambda_0$, states with
two dipoles on a given link or on two consecutive links do not form
a part of the low-energy subspace \cite{subir1}. Thus the effective
dipole model describing the low-energy physics of the model can be
written as \cite{subir1}
\begin{eqnarray}
H_d &=& \sum_{\ell} (-w (d_{\ell}^{\dagger} + d_{\ell}) \,
+\mu_d(\ell)
\hat n^d_{\ell}) \nonumber\\
&& n^d_{\ell} \le 1, \quad n_{\ell}^d n_{\ell +1}^d =0
\label{dipoleham1}
\end{eqnarray}
where $\hat n_{\ell}^d = d_{\ell}^{\dagger} d_{\ell}$ is the dipole
number operator and $w= J \sqrt{n_0(n_0+1)}$. The constraints
$n^d_{\ell} \le 1$ and $n_{\ell}^d n_{\ell +1}^d =0$ truncate the
size of the Hilbert space by eliminating states with dipoles on
adjacent sites. The ground state phase diagram of the model, which
contains a $Z_2$ symmetry broken phase as shown in Fig.\
\ref{fig1}(b), is charted out in the supplemental
information~\cite{supp1}.

\begin{figure}[t!]
\begin{center}
\includegraphics[width=\columnwidth]{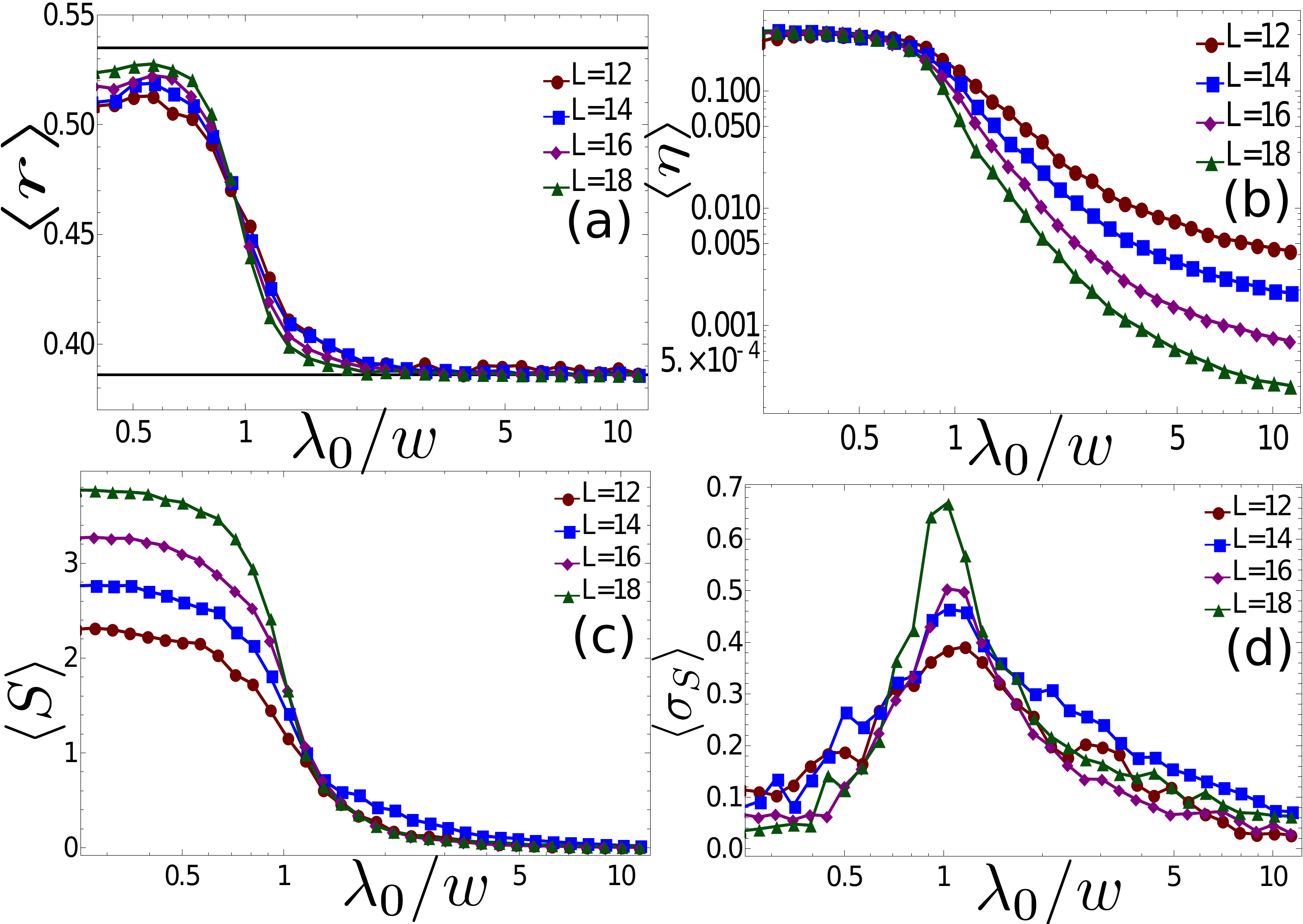}
\end{center}
\caption{(a) Plot of $\langle r\rangle$ as a function of $\lambda_0$
showing an ergodic-MBL transition at $\lambda_0/w \simeq 0.946$. (b)
A plot of NPR, $\eta$ as a function of $\lambda_0$ for several $L$.
(c) Plot of the half-chain entanglement entropy $S$ as a function of
$\lambda_0$. (d) Plot of fluctuation of entanglement, $\sigma_s$, as
a function of $\lambda_0$. For all plots, $U_0-{\mathcal E}/w=0$. }
\label{fig3}
\end{figure}

We now focus on the ergodic-MBL transition in this model. At the
outset, we note that a MBL phase of the dipole model (Eq.\
\ref{dipoleham1}) does not amount to that of the original boson
model (Eq.\ \ref{ham0}) since the latter has additional states which
are not part of the former's Hilbert space. However, we expect
experiments discussed in Ref.\ \onlinecite{greiner2} (which emulates
$H_d$ (Eq.\ \ref{dipoleham1}) using a Rydberg atom chain) to exhibit
the MBL phase of $H_d$. We shall discuss this point later in
details. To study the MBL phase and the associated ergodic-MBL
transition, we fix ${\mathcal E}$ and obtain the eigenvalues and
eigenvectors of the model for $L\le 18$ using ED for several values
of $\lambda_0$. We then use these to compute three quantities which
may distinguish between ergodic and MBL phases.

The first of these is the ratio of the difference of successive gaps
in the energy spectrum, $r_n = {\rm Min}
[\Delta_{n+1}-\Delta_{n}]/{\rm Max[\Delta_{n+1}-\Delta_n]}$, where
$\Delta_n= E_{n+1}-E_{n}$, and $E_n$ denotes the eigenvalues of
$H_d$. It is well known that $r_n$ obeys Poisson (GOE) statistic in
the MBL (ergodic) phase with the mean value $\langle r\rangle =0.386
(0.535)$ \cite{mblrev1,mblrev2}. The second, is the NPR defined as
$\eta = \sum_n |\psi_n|^4/{\mathcal D}$, where $\psi_n = \langle
n|\psi\rangle$, $|\psi\rangle$ is the wavefunction of a typical
state with finite energy density, and $|n\rangle$ denotes
eigenstates of $\hat n_d$. $\eta$ is expected to be a system-size
independent constant in the ergodic phase; in contrast, it decays
exponentially with system size in the MBL
phase\cite{mblpapersanalytical}. Finally, we compute the
entanglement entropy $S= -{\rm Tr} \rho \ln \rho$ for a given
subsystem of length $L/2$ described by a density matrix $\rho$ for a
representative state with finite energy density in the middle of the
spectrum. For such a typical state, $S$ follows a volume(area) law
in the ergodic(MBL) phase \cite{mblrev1,mblrev2}. In addition, we
also compute the fluctuation of the entanglement entropy,
$\sigma_S$, as a function of $\lambda_0$. Each of these quantities,
as we find below, provides an independent measure to discern between
ergodic and MBL phases.

The results obtained from these calculations are shown in Fig.\
\ref{fig3} which indicate a finite-size crossover from ergodic to
MBL phase around $\lambda_0 \simeq w$. Fig.\ \ref{fig3}(a) shows
that $\langle r\rangle$ changes from its expected values in the
ergodic phase to that in the MBL phase around $\lambda_{0c} \simeq
0.95 w$ where curves corresponding to different $L$s cross. A
similar trend is noticed in Fig.\ \ref{fig3}(b) where $\eta$ becomes
$L$ dependent around $\lambda_{0c}$ indicating the onset of a MBL
phase. In Fig.\ \ref{fig3}(c), we find that $S$ becomes independent
of system size for $\lambda_0 >\lambda'_{0c} \simeq 1.25 w$
indicating an area law behavior for a generic state in the middle of
the spectrum and therefore a MBL phase. Finally, in Fig.\
\ref{fig3}(d), we find enhancement of $\sigma_s$ around
$\lambda'_{0c}$ indicating presence of strong quantum fluctuation at
this point. Thus our data establishes the presence of a MBL phase in
the constrained dipole boson model. Note that the critical
$\lambda_{0}$ from the two diagnostics ($\langle r\rangle$ and $S$)
do not in general agree exactly for system sizes accessible to ED
~\cite{naldesi}.

\begin{figure}[t!]
\begin{center}
\includegraphics[width=\columnwidth]{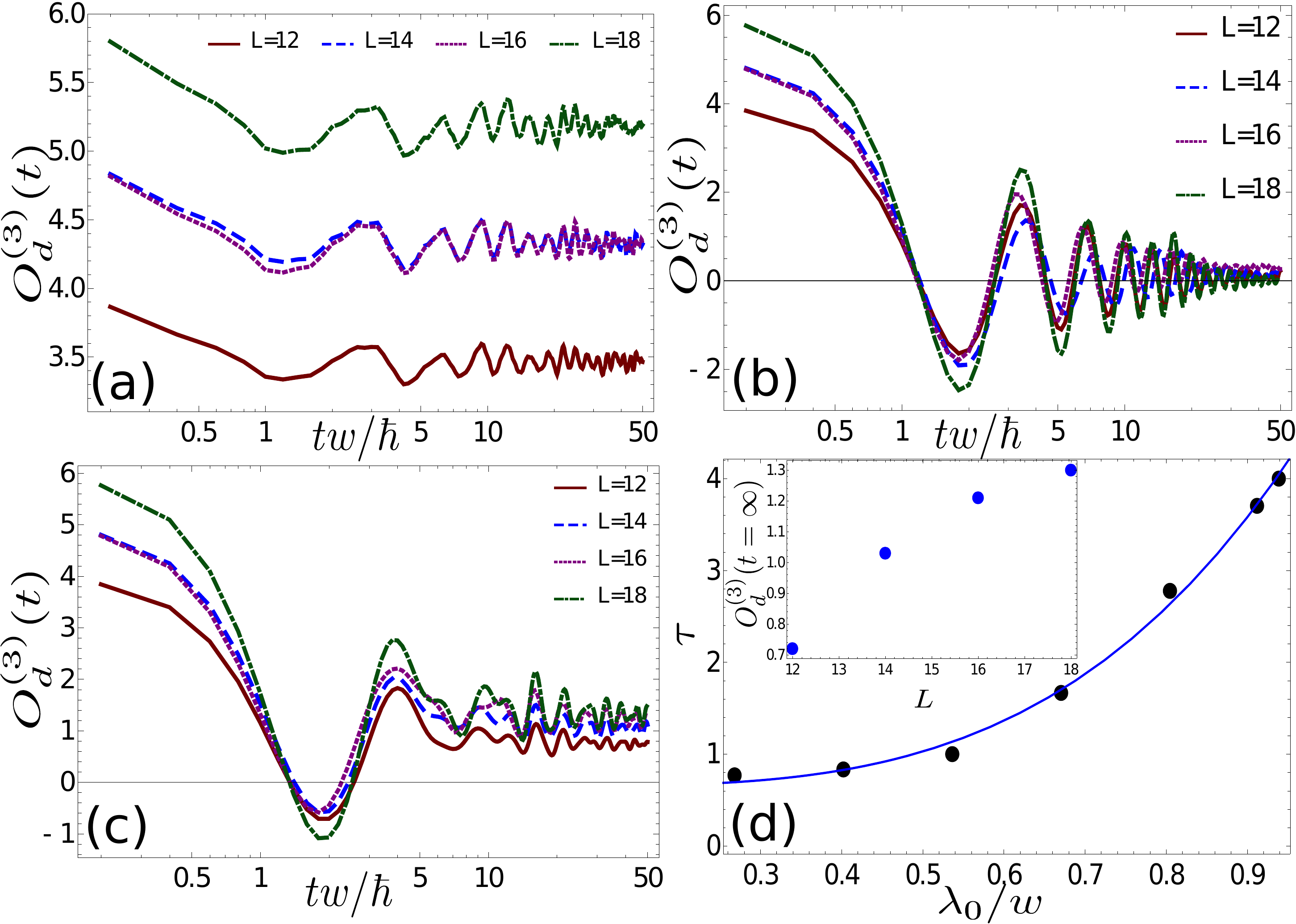}
\end{center}
\caption{(a) Plot of  $\langle O_d^{(3)}\rangle(t)$ as a function of
time in the MBL phase with $\lambda_0/w=5.37$. (b) Same as in (a)
but in the ergodic phase with $\lambda_0/w =0.268$. (c) Same as in
(a) at $\lambda_{0c}=0.95 w$. (d) Plot of the decay time $\tau$ of
the oscillations of $O_d^{(3)}(t)$ as a function of $\lambda/w$. The
inset shows the $L$ dependence of $O_d^{(3)}(t \to \infty)$ at
$\lambda_{0c}$. See text for details.} \label{fig4}
\end{figure}

The experimental signature of such a MBL phase is most easily picked
up in dynamics. To this end, we study the behavior of the dipole
order parameter in the $Z_3$ symmetry broken ground state, $\langle
O_{d}^{(3)}\rangle(t)$, as a function of time following a sudden
quench of $\lambda_0$ (for behavior of $S$ following such a quench
see Ref.\ \onlinecite{supp1}). We start from the $Z_3$
symmetry-broken dipole-ordered state $|\psi_0(t=0)\rangle$, (which
is a ground state of $H_{\rm Ryd}$ as discussed above), and perform
a sudden quench of $V_{jj'}$ $\lambda_0$ with $\Delta \equiv
(U-{\mathcal E}_0)/w=5$ so that the final Hamiltonian has a $Z_2$
symmetry broken ground state. One can then write $|\psi(t)\rangle =
\sum_{c_m} \exp[-i E_m t/\hbar] |m\rangle$, $|m\rangle$ and $E_m$
denote the eigenstates and eigenvalues of the new Hamiltonian and
$c_m = \langle m|\psi_0(t=0)\rangle$. Using this one obtains
\begin{eqnarray}
\langle O_d^{(3)}\rangle (t) &=& \sum_{m,n} c_m^{\ast} c_n e^{-i
(E_n-E_m)t/\hbar} \langle m|O_d^{(3)}|n\rangle.
\end{eqnarray}

A plot of $\langle O_d^{(3)}\rangle (t)$ as a function of time,
shown in Fig.\ \ref{fig4}(a) clearly shows that in the MBL phase
where $\lambda_0 \gg \lambda_{0c}$, $\langle O_d^{(3)}\rangle (t)$
remains finite over a long period of time. In contrast, as shown in
Fig.\ \ref{fig4}(b), in the ergodic phase where $\lambda_0 \ll
\lambda_{0c}$, it decays to zero over a short time scale after a few
oscillations. At $\lambda_0 =\lambda_{0c}$, as shown in Fig.\
\ref{fig4}(c), $\langle O_d^{(3)}\rangle (t)$ oscillates initially
but decays to a final value which approaches zero for $L \to \infty$
as shown in the inset of Fig.\ \ref{fig4}(d). The decay time $\tau$
of $O_d^{(3)}(t)$ is obtained by fitting its oscillation envelope to
$A \exp[-t/\tau]$ where $A$ and $\tau$ are fitting parameters. In
the ergodic phase, as shown in Fig.\ \ref{fig4}(d), $\tau$ increases
with $\lambda$ with $\tau w/\hbar \simeq 4$ for $\lambda=0.94$; it
diverges in the MBL phase.

The most suitable platform for experimental realization of our work
constitutes an array of Rydberg atoms. These systems, in the absence
of AA potential, have recently been realized experimentally in Ref.\
\onlinecite{greiner2}. The Hamiltonian of these Rydberg atoms is
given by
\begin{eqnarray}
H_{\rm Ryd} &=& \sum_j (-\Omega \sigma_j^x + \Delta_0 \hat n_j) +
\sum_{jj'} V_{jj'} \hat n_j \hat n_{j'} \label{rydham1}
\end{eqnarray}
where $\hat n_j$ denotes the number operator for Rydberg (excited)
atoms on site $j$, $\Delta_0$ denotes detuning parameter which can
be used to excite an atom to a Rydberg state, $V_{jj'} \sim
1/|x_j-x_{j'}|^6$ denotes the interaction strength between two
Rydberg atoms and $ \sigma_j^x = |r_j\rangle \langle g_j| +
|g_j\rangle \langle r_j|$ denotes the coupling between the Rydberg
($|r_j\rangle$) and ground ($|g_j\rangle$) states. We note that for
$V_{jj'}=0$, Eq.\ \ref{rydham1} can be directly mapped to Eq.\
\ref{dipoleham1} via the identification $\Omega \to w$, $\Delta_0
\to \mu_d$ and $\hat n_j \to \hat n_d$. In experiments, $V_{jj'}$
could be tuned so that $V_{j \,j+1} \gg \Delta_0, \Omega$ and $V_{j
\,j+n} \ll \Delta, \Omega$ for $n>1$. This effectively implements
the constraint ${\hat n^d}_{\ell} {\hat n^d}_{\ell+1}=0$ leading to
realization of $H_d$ with $Z_2$ symmetry broken ground state for
$\Delta_0 \ll 0$ \cite{greiner2}. Other configurations of $V_{jj'}$
where $V_{j \,j+n} \gg \Delta_0, \Omega$ for $n=1,2$ led to
experimental realization of the $Z_3$ states. Such a state turns out
to also be the ground state of $H_d$ supplemented with an additional
dipole-dipole interaction term \cite{subir3,rg1}. We also note that
there have been concrete proposals for realization of the AA
potential for ultracold atom chains \cite{aaref1}. In what follows,
we propose that such potentials are applied on the Rydberg atom
chain studied in Ref.\ \onlinecite{greiner2}.

The dynamics of $\langle O_d^{(3)}\rangle (t)$ can be studied
experimentally by first preparing a Rydberg chain in a ground state
of $H_{\rm ryd}$ with $V_{j \,j+n} \gg \Delta_0, \Omega$ for $n=1,2$
and $\lambda_0=0$. This is to be followed by sudden quenches of
$V_{jj'}$ and $\lambda_0$ such that $V_{j \,j+n} \gg \Delta_0,
\Omega$ for $n=1$ and $\lambda_0$ has a desired finite value. Such
quenches can be experimentally performed by tuning suitable laser
intensities \cite{greiner2}. Our prediction regarding post-quench
dynamics of $O_d^{(3)}(t)$ is as follows. Below
$\lambda_0=\lambda_{0c}$, $O_d^{(3)}(t)$ will decay to zero with a
characteristic timescale $\tau$ signifying the ergodic phase. The
value of $\tau$ will diverge at $\lambda_{0c}$. For $\lambda_0
> \lambda_{0c}$, $O_d^{(3)}(t)$ will remain close to its original
value for $ t \gg \tau(\lambda_0)$. We note that $O_3^{(d)}(t)$ can
be easily obtained for the present model via measurement of $\langle
\hat n_j \rangle$. In earlier experiments, this was achieved by
measuring parity of occupation of the Rydberg
atoms\cite{greiner1,greiner2}.

Our proposal provides an estimate on the lower bound of timescale
over which $O_d^{(3)}(t)$ needs to remain finite for claiming
experimental realization of the MBL phase. The lifetime of a Rydberg
chain is primarily determined by atom loss from the trap and is
typically around $10 \mu$s for realistic experimental
parameters\cite{greiner2}. From Fig.\ \ref{fig4}(d), we find that
the maximal decay time in the ergodic phase near the transition is
$\tau_{\rm max} \simeq 4 \hbar/w \equiv 4 \hbar/\Omega$. Thus for
$\Omega= 4 \pi$MHz, one needs to follow the dynamics for $T\gg
\tau_{\rm max}\simeq 0.3 \mu$s. This requirement can be met since
typical experimental timescale $t_{\rm expt} \sim 7\mu s \simeq 22
\tau_{\rm max}$ \cite{greiner2}. From Fig.\ \ref{fig4}, we indeed
find that $O_d^{(3)}(t)$ decays close to zero for $t \simeq 10
\tau_{\rm max}= 40\hbar/w \simeq 3\mu$s in the ergodic phase while
it retains a finite value after this time in the MBL phase.
\begin{figure}[t]
    \begin{center}
        \includegraphics[width=0.49\columnwidth]{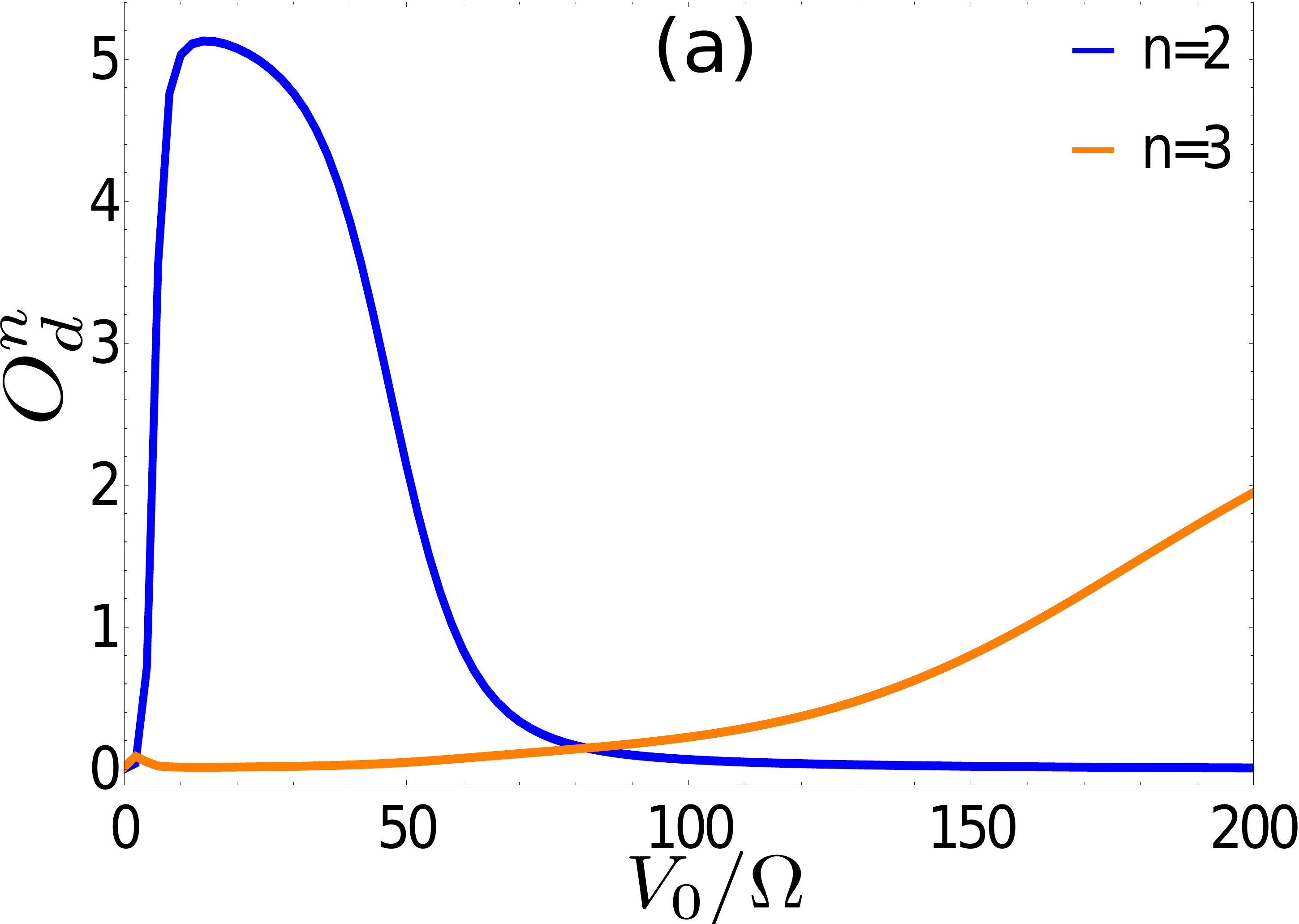}
        \includegraphics[width=0.49\columnwidth]{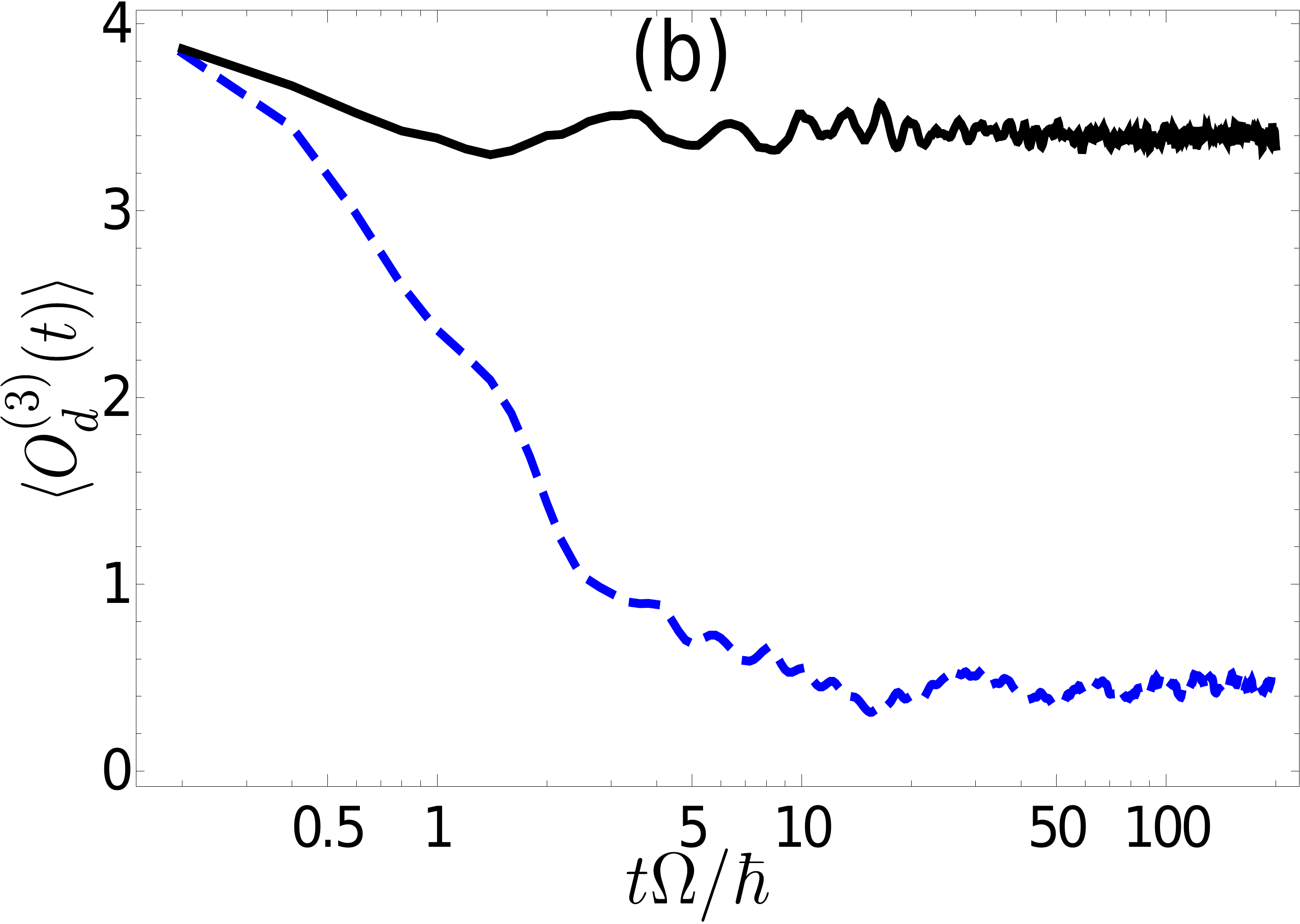}
    \end{center}
    \caption{(a) A plot of ground state value of the $Z_2$ and $Z_3$ order parameters as a function of $V_0/\Omega$
    for a $L=12$ Rydberg chain. We have chosen $\Delta_0/\Omega=-3$. (b) Plot of $O_d^{(3)}$
    in the ergodic ($\lambda_0/\Omega=0.5$) and MBL ($\lambda_0/\Omega=10$)
    phases for $V_0/\Omega=40$, $\Delta_0/\Omega=-3$. }
    \label{S3}
\end{figure}

Finally, we provide explicit numerical evidence for signature of the
MBL phase in the Rydberg atom chain using ED. An estimate of $V_0$
may be obtained from an ED study of the ground state $H_{\rm ryd}$
for $L=12$. We find that for $\Delta_0/\Omega=-3$ the $Z_2$ symmetry
broken order occurs for $V_0/\Omega \le 80$ (Fig.\ \ref{S3}). Next
we turn on the the AA potential $\lambda_j$ and study the response
of the dipole order parameter $O_d^{(3)}$ for the Rydberg atoms. We
note that the present system does not appear to exhibit a mobility
edge; all states are either ergodic or localized depending on the
value of $\lambda_0$. We also find that $O_d^{(3)}$ exhibits
ergodic(MBL) behavior for small (large) $\lambda_0$ (Fig.\
\ref{S3}). The value of $\lambda_{0c}$ and the period of oscillation
of $O_d^{(3)}$ in the ergodic phase is found to approach those
obtained from analysis $H_d$ with increasing $V_0$ as expected. We
thus conclude that a choice of large $V_0$ which is well within
acceptable experimental parameter range would allow one to study the
MBL phase of $H_d$ using $H_{\rm Ryd}$.

In conclusion, we have obtained the phase diagram and demonstrated
the existence of an ergodic-MBL quantum phase transition for a
dipole boson model in the presence of the AA potential. We have also
shown that the non-equilibrium dynamics model picks up signatures of
the MBL phase of this model, discussed the relevant timescales
involved, and suggested concrete experiments using Rydberg atom
chains which can test our theory.\\

{\it Acknowledgement}: A.D. acknowledges funding from SERB NPDF
research grant PDF/2016/001482. SM acknowledges funding from the UGC
through the Indo-Israeli project.

\appendix

\section{Supplemental Material for Many-body localized phase of
bosonic dipoles in a tilted optical lattice}

\subsection{Ground state phase diagram}

\begin{figure}[h!]
\begin{center}
\includegraphics[width=0.49\columnwidth]{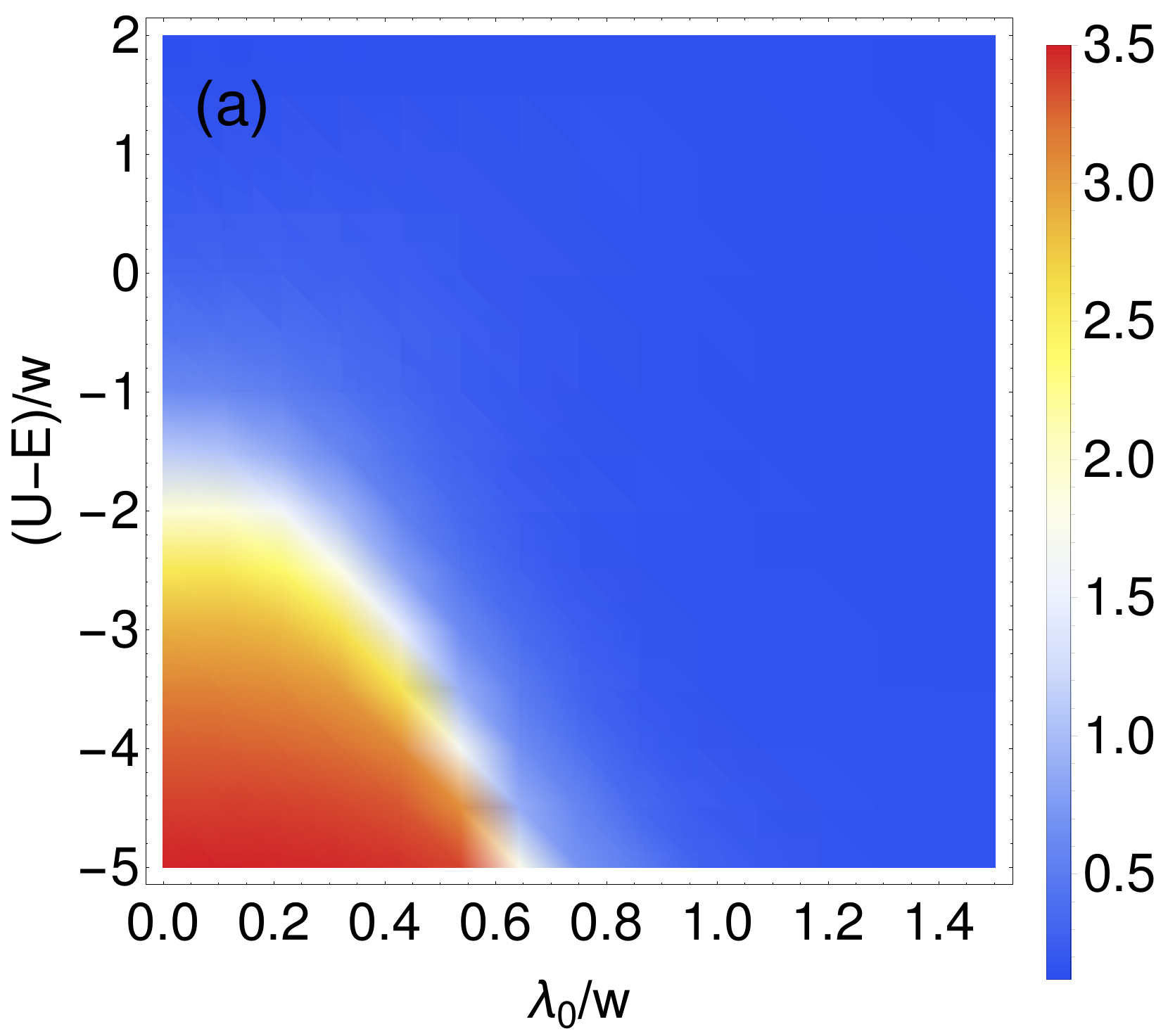}
\includegraphics[width=0.49\columnwidth]{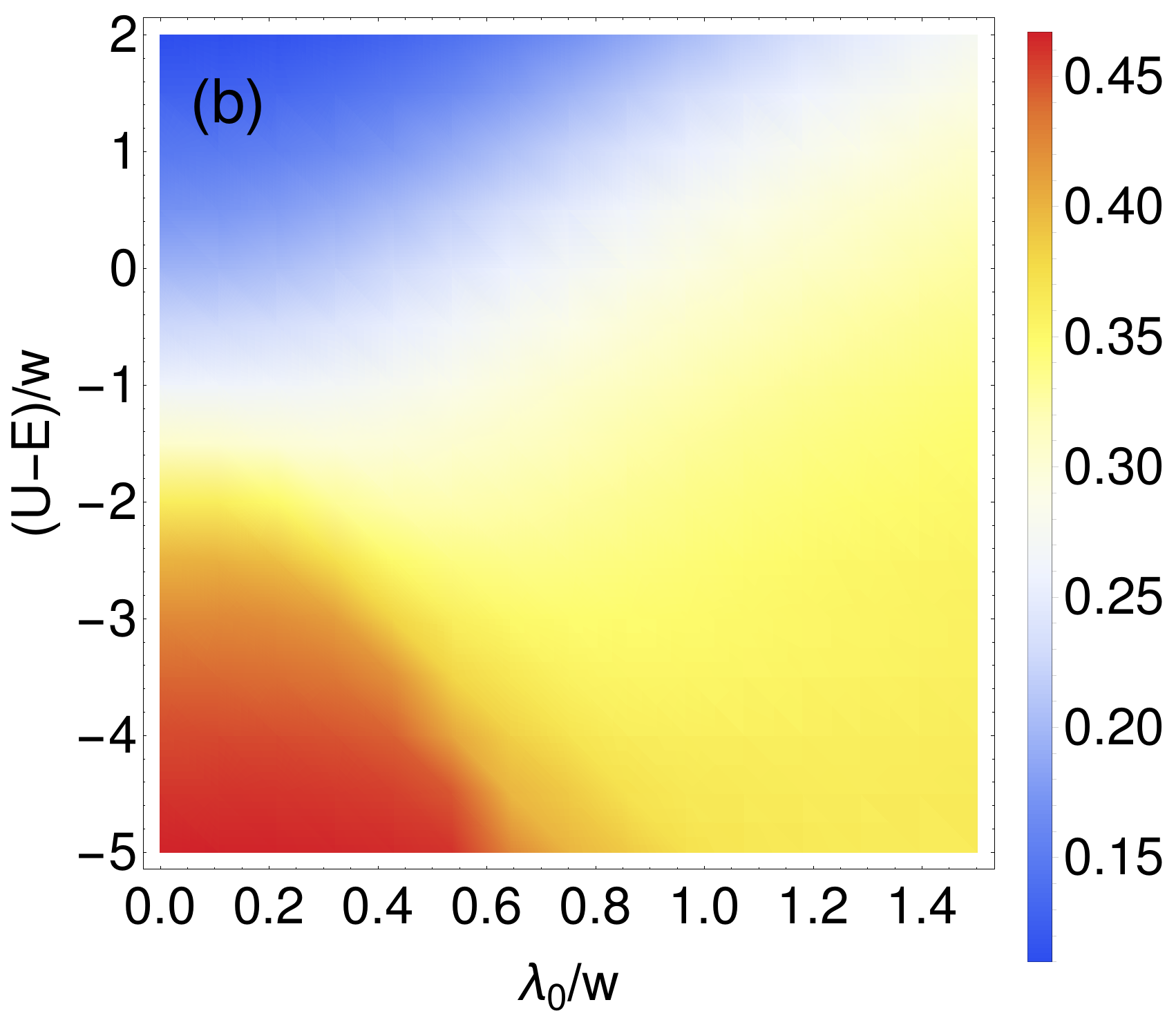}
\end{center}
\caption{(a) Phase diagram of the dipole model with AA potential
showing the plot of the dipole order parameter $O_d^{(2)}$ as a
function of $(U-{\mathcal E})/w$ and $\lambda_0/w$ for $L=18$. (b)
The dipole density $n^d$ as a function of $(U-{\mathcal E})/w$ and
$\lambda_0/w$.} \label{fig2}
\end{figure}

The ground state phase diagram of the model, as obtained by ED for
$L=18$ is shown in Fig.\ \ref{fig2}(a). To chart out the phase
diagram, we plot the dipole order parameter $O_d^{(2)}$ in the $Z_2$
symmetry broken phase as a function of $(U-{\mathcal E})/w$ and
$\lambda_0/w$. As expected, we find that for any given ${\mathcal
E}$, an increase in $\lambda_0$ decreases the magnitude of
$O_d^{(2)}$ and eventually the ordered phase is destroyed via a
melting transition to a disordered phase. Similarly, for any given
$\lambda_0$, increasing ${\mathcal E}$ increases the tendency
towards an ordered phase. We note here that the presence of the AA
potential may either aid or hinder dipole formation on a given link
depending on the value of $\sin(2 \pi \beta \ell)$. This can be
clearly seen from Fig.\ \ref{fig2}(b) where $n^d = \langle
\sum_{\ell} \hat n^d_{\ell} \rangle/L$ is plotted as a function of
$\lambda_0/w$ and $(U-{\mathcal E})/w$ showing an increase of $n_d$
with increasing $\lambda_0$ for any ${\mathcal E}$. However,
$O_d^{(2)}$ always decreases with increasing $\lambda_0$.

\subsection{Growth of the entanglement entropy in dynamics}

In the main text, we showed that the entanglement entropy $S$ allows
us to distinguish between the many-body localized phase and the
ergodic phase. The growth of entanglement following a quench is also
a key notion in understanding the ergodic-MBL transition. The
entanglement entropy following a quench grows linearly in time
ergodic phase. In contrast, its growth is logarithmic in the MBL
phase. In both cases, the entanglement entropy eventually saturates
to a value, that scales with system size $L$.

\begin{figure}[h!]
    \begin{center}
        \includegraphics[width=\columnwidth]{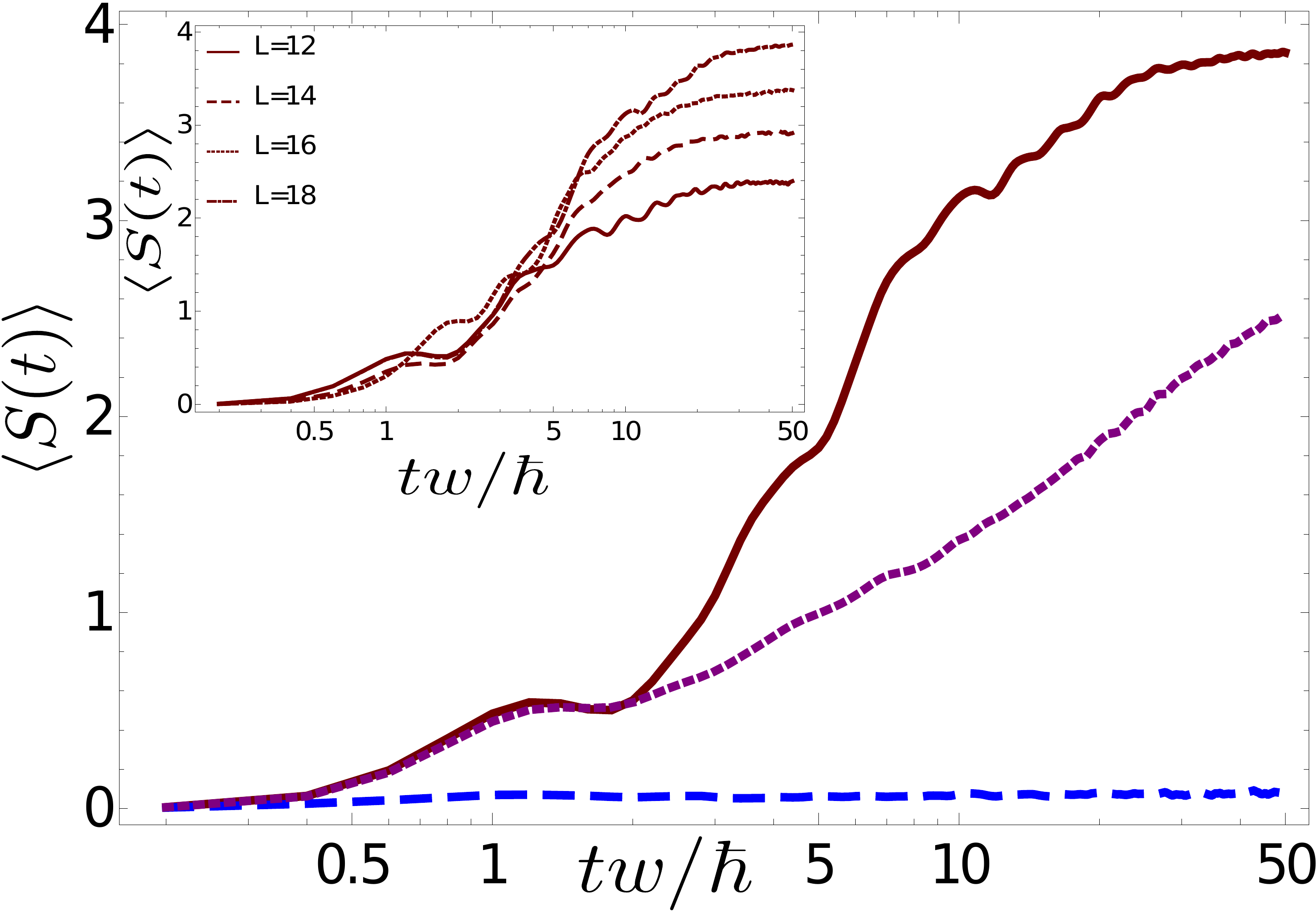}
    \end{center}
    \caption{The entanglement entropy after a quench
    with time for system size $L=18$. The brown
    line is for a quench to the ergodic phase with
    $\delta \lambda/w=0.5$. The blue line for a quench
    to the MBL phase with $\delta \lambda/w=20$. The
    purple lines are for quench close to the critical
    point with $\delta \lambda/w=3.095$. The inset shows value at
    saturation depends on the system size.}
    \label{S1}
\end{figure}

To study the behavior of $S$, we follow the same quench protocol as
in the main text. We evaluate the half-chain entanglement entropy
(with subsystem size $L/2$, where $L$ is the chain length) as a
function of time. The result is plotted in Fig.\ \ref{S1}. In the
ergodic regime, as shown in Fig.\ \ref{S1}, $S(t)$ grows linearly
and saturates to a value proportional to the system size $L$. The
inset shows the system size dependence of entanglement entropy in
the ergodic phase. In contrast, in the MBL phase, the growth
logarithmic in time while close to the critical point the growth is
faster than that inside the MBL phase. As we increase $\lambda$, we
clearly find a transition from linear to logarithmic behavior
marking a transition (crossover for finite size) from an ergodic to
an MBL phase. We find that $S$ is independent of the system size in
the MBL phase while its $L$ dependence in the ergodic phase is shown
in the inset of Fig.\ \ref{S1}.

\subsection{Bosonic dipoles in disordered and tilted optical lattice}

The 1D dipole Hamiltonian we consider in Eq.\ 3 of the main text
undergoes an ergodic-MBL transition upon varying the strength of the
quasiperiodic potential $\lambda_0$. However, the more standard
setting in which MBL is observed involves the presence of quenched
disorder and not a quasiperiodic potential. Here we show that the
constrained dipole model in the presence of quenched disorder also
undergoes and ergodic-MBL transition. We begin by, switching off the
quasiperiodic potential by setting $\lambda_0$ to zero, in the
dipole Hamiltonian, Eq.\ (3) and instead introducing an on-site
disorder potential term,
\begin{eqnarray}
H_d &=& \sum_{\ell} (-w (d_{\ell}^{\dagger} + d_{\ell}) \,
+\mu_d(\ell)
\hat n^d_{\ell}) \nonumber\\
&& n^d_{\ell} \le 1, \quad n_{\ell}^d n_{\ell +1}^d =0
\label{dipoleham2}
\end{eqnarray}
where $\hat n_{\ell}^d = d_{\ell}^{\dagger} d_{\ell}$ is the dipole
number operator and
\begin{eqnarray}
\mu_d(\ell)&=&\mu_0(\ell)+\delta \mu_{\ell}
\end{eqnarray}
\begin{figure}[t]
    \begin{center}
        \includegraphics[width=\columnwidth]{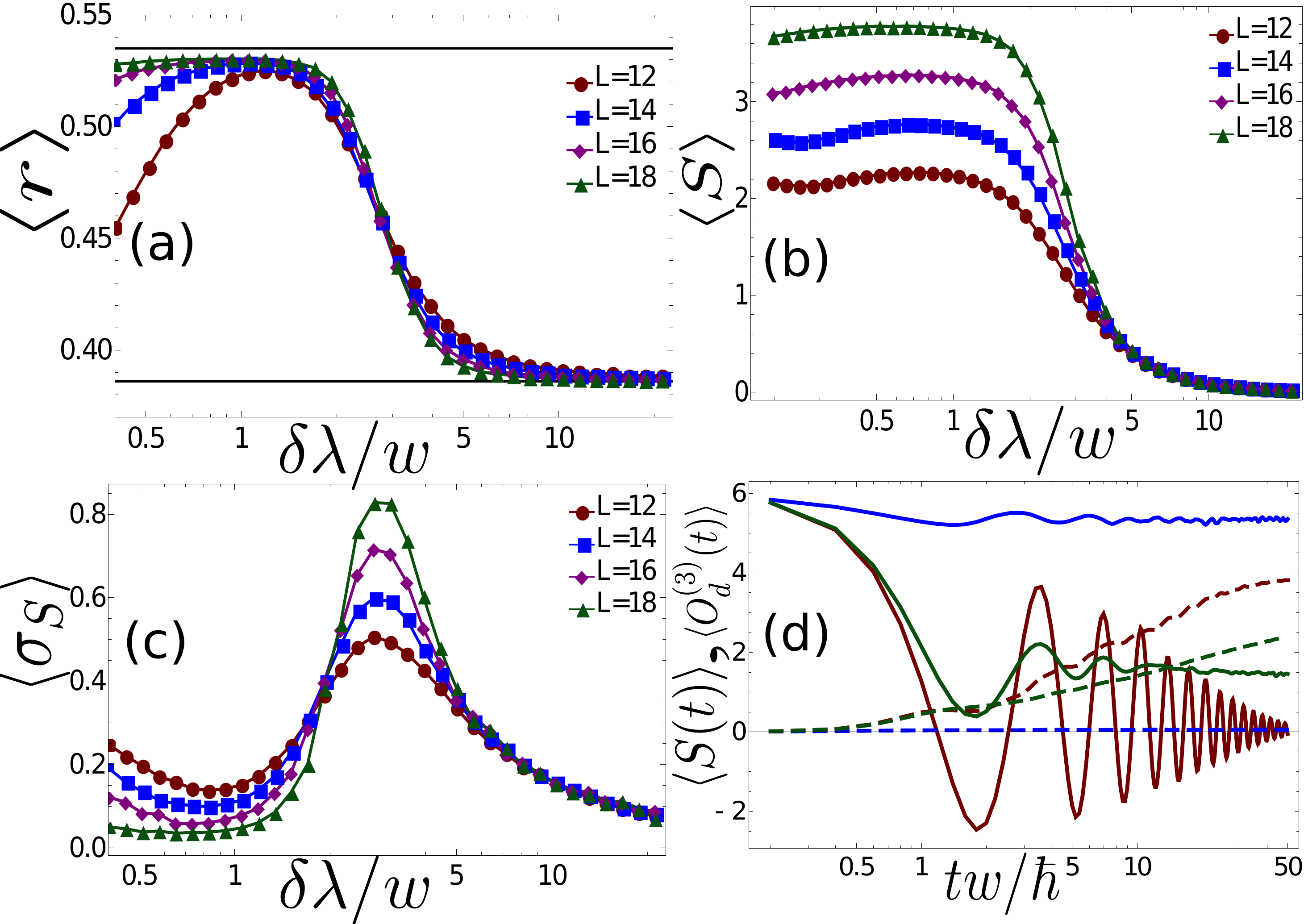}
    \end{center}
    \caption{(a)$\langle r \rangle $ as a function of $\delta\lambda$
showing an ergodic-MBL transition at 3.095. (b) The half-chain
entanglement entropy(S) with $\delta\lambda$. (c) The fluctuation of
entanglement entropy $\sigma_S$ with $\delta\lambda$. (d)Plot of
$\langle O_d^{(3)}(t)\rangle$(solid lines) and entanglement entropy
(dashed lines) after a quench as a function of time for system size
$L=18$. The red line is for a quench to the ergodic phase, blue for
a quench to the MBL phase and green for a quench close close to the
ergodic-MBL critical point.}
    \label{S2}
\end{figure}
where $\mu_0(\ell)=U-\mathcal{E}$ and $\delta \mu_{\ell}$ are
independent random variables drawn from a uniform distribution
$\left[-\delta\lambda,\delta\lambda\right]$. When $\delta
\lambda=0$, the model reduces to the dipole model studied earlier
and in the presence of weak disorder the Hamiltonian is
ergodic\cite{subir1}. At a finite critical disorder strength, we
show the system displays an ergodic-MBL transition. We use the same
quantities discussed in the main text to understand the transition.
The results obtained are shown in Fig:\ref{S2}. First we plot in
Fig\ref{S2}(a) the difference of the successive gaps in the energy
spectrum with the strength of the disorder which shows a transition
from ergodic to MBL around the critical value of
$\delta\lambda_c/w=3.09$. In Fig \ref{S2}(b) we compute the
entanglement entropy for a typical state at finite energy density in
the middle of the spectrum as a function of the strength of the
disorder, which also shows a transition from the ergodic to MBL
phase. We have also computed the fluctuations in entanglement
entropy which too indicates a transition from the ergodic to MBL
phase with large fluctuations at the critical point. Finally in
Fig.\ \ref{S2} we study the nonequilibrium dynamics of the system
with the same protocol discussed in the main text. We have plotted
the order parameter $\langle O_d^{(3)}(t)\rangle$(solid line) and
entanglement entropy(dashed line) of the time evolved state for
system size $L=18$ for three different regimes. The red line for a
quench to the ergodic phase, the blue line for a quench to the MBL
phase and the green line for a quench close to critical point. In
the ergodic phase, the order parameter decays to zero and
entanglement entropy grows linearly to a saturation value
proportional to system size $L$ over a short time scale. In the MBL
phase, the order parameter remains close to the initial value and
the entanglement entropy increases very slowly (logarithmically) in
time. Close to the critical point the order parameter decays slowly
and the entanglement entropy grows slowly compared to ergodic phase.

\end{document}